# Size-dependent failure behavior of commercially available lithium-iron phosphate battery under mechanical abuse

Vishesh Shukla[a], Ashutosh Mishra[a*], Jagadeesh Sure[b], Subrata Ghosh[c], R.P. Tewari[a]

[a]Department of Applied Mechanics, Motilal Nehru National Institute of Technology Allahabad, Prayagraj, Uttar Pradesh211004, India
[b]Department of Physics, School of Advanced Sciences, Vellore Institute of Technology (VIT), Vellore, Tamil Nadu632014, India
[c]Micro and Nanostructured Materials Laboratory-Nano Lab, Department of Energy, Politecnico di Milano20133, Milano, Italy

*Corresponding author e-mail: amishra@mnnit.ac.in; ashutoshjssate@gmail.com

**Abstract**

Under mechanical abuse conditions, failure of lithium-ion batteries occurs in stages characterized by different force, temperature and voltage response which require it's in situ measurements for analysis. Firstly, four sizes of commercially available lithium-iron phosphate batteries (LFPB) viz. 18650, 22650, 26650, and 32650 are subjected to quasi static lateral, longitudinal compression, and nail penetration tests. The failure, characterized by the voltage drop and temperature rise, at the onset of the first internal short-circuit (ISC), is identified by Aurdino-based voltage sensor module and temperature measurement module. The battery failure load and peak temperature at the onset of ISC are found to rely on the battery size strongly. The failure is observed to be delayed for small-sized 18650 batteries during lateral compression, unlike longitudinal compression and nail penetration test. At the onset of the short circuit, the temperature rise above ambient value is different for different LFPBs. It is found to be maximum (64.4 °C) for LFPB 32650 under longitudinal compression and minimum (29.5 °C) under lateral compression tests amongst the considered geometries. Further, LFPB 26650 exhibited a balanced thermal behavior during the considered abused condition. Such data can be sensed timely for effective thermal management and improved safety of lithium-ion batteries.

**Keywords:** Lithium-ion battery, battery failure, Lithium-iron phosphate battery, Mechanical integrity, Mechanical abuse, Internal short circuit, Compression loading

## Abbreviations and nomenclature

| | | | | |
|---|---|---|---|---|
| LFPB | : | Lithium-iron phosphate battery | BTMS | : Battery Thermal Management System |
| EV | : | Electric vehicle | PNE | : Potential of Negative Electrode |
| LIB | : | Lithium-ion battery | PCM | : Phase Change Material |
| IFR | : | International Federation of Robotics | UBHC | : Upper Bound of Heating Current |
| LCOB | : | Lithium cobalt oxide batteries | mAh | : Mili-Ampere hour |
| LMOB or NMC | : | Lithium manganese oxide batteries | Ah | : Ampere hour |
| ISC | : | Internal short circuit | °C | : Degree Celsius |
| SOC | : | States of charge | mm | : Millimeter |
| PTC | : | Positive temperature coefficient | G | : Gram |
| LFP | : | Lithium Iron phosphate | V | : Volts |
| AIS | : | Automotive Industry Standards | Cm | : Centimeter |
| UL | : | Underwriters Laboratories | $mm^3$ | : cubic millimeter |
| USABC | : | United States Advanced Battery Consortium | Wh/kg | : Watt hour per kilogram |
| CAR | : | Cooperative Automotive Research | mg/Wh | : milligram per Watt hour |
| UN/ECE | : | United Nations Economic Commission for Europe | kN | : Kilo-Newton |
| GB/T | : | Guobiao standards | | |

# 1.   Introduction

Use of batteries is increasing day by day in our daily life, starting from the alarm clock, laptops to smart phones, electric vehicles (EVs), and so on. Moreover, a rapid switch over to battery-based clean transportation technologies from oil-based transportation is one of the probable solutions to mitigate the air pollution. In most cases, the use of lithium-ion batteries (LIBs) is already commercialized due to their high energy density. However, a greater tendency to catch fire and explode, when subjected to undesired loading such as in case of EVs crash or overheating due to beyond design basis operating conditions, is very dangerous and unlikely. Growing cases of fire in EVs, due to battery failure [1-4] have raised an important safety concern among the EV manufacturers towards developing stringent safety standards for EV batteries such as GB/T 31485–2015 developed by China [5,6]. Likewise, UL 2580, USABC and Freedom CAR of USA, KMVSS 18-3 of Korea, UN/ECE-R100.02 of the European Union, and AIS-048 of India, have been developed [7]. However, the dynamics of the failure behavior of lithium-ion batteries are still not thoroughly understood. Due to this reason, safety assessments of batteries are much needed for time-to-time revisions in the safety standards.

According to recent evidence [8], LFPBs have a safer operation compared to lithium cobalt oxide batteries (LCOBs) and lithium manganese oxide batteries (LMOBs or NMCs). Being advantageous due to greater thermal stability and crash safety, LFPBs are gaining the attention of EV manufacturers. Therefore, in day, significant research activities are seen focusing on the safety testing of batteries under mechanical and thermal abuse loadings [9]. Commonly, mechanical abuse tests involve the prediction of failure load under different types of quasi-static and dynamic load conditions [10,11]. Besides, loss of mechanical integrity leading to thermal runaway in a battery can become catastrophic, hence its risk mitigation strategies require thorough understanding of battery failures. Therefore, tests that discuss the mechanical abuse-driven thermal runaway behavior become important. In this context, Hatchard *et al.* [12] developed a "smart nail" for collecting temperature *versus* time data at the point of nail penetration. This nail with an attached thermocouple on the cell surface and tabs were used to measure voltage. The tests provided new insights into the behavior of the LIB during nail penetration conditions. Fang *et al.* [13] performed an ISC test of 1000 mAh LIB. They observed a temperature spike in the *anode-aluminum short*, in comparison to that occurred at the *anode-cathode short*. Because ISC triggers the thermal runaway, the exothermic chemical reactions may result in fire and explosion if the onset of failure is not detected timely. In this direction, other researchers [14,15] have attempted to characterize the failure due to ISC that occurs layer by layer with voltage drop and repeated partial recovery. A series of indentation tests were conducted till the occurrence of ISC by Zhu *et al.* [16] for different capacities of LIBs. It was pointed out that the cells with different capacities and states of charge exhibited different behaviors. Wang *et al.* [17] analyzed the mechanism of nail penetration-induced ISC of a lithium-ion cell. The results show that the nail plays a dual role during the penetration process causing a rupture in the casing and thermal runaway due to short-circuiting between electrodes and subsequent fume/electrolyte release. During the rupture process due to the nail penetration test, there were two modes of temperature rise in case of thermal runaway condition. It was found that, for the two modes, the time involved in the penetration to cause thermal runaway, was 5 sec and 60 sec respectively.

Apart from the effect of different loading conditions, investigations on the role of form factors and diametral variations of cylindrical batteries on the failure have been also conducted. Zeng *et al.* [18] discussed the effect of cathode materials on the hazard level of LIB with 18650 form-factor and revealed the significance of choice of active materials. It was observed that the thermal runaway time of LCOBs is shorter than NMCs and LFPBs. Kim *et al.* [19] developed a 3D model to simulate the oven tests of 18650 and 50900 LCOBs at 150 °C and 155 °C. The results indicated that the thermal runaway did not occur in 18650 batteries but did occur in 50900 batteries indicating the profound effect of battery size on thermal runaway behavior. Duh *et al.* [20] investigated the thermal runaway hazards for cylindrical LFPBs of different sizes (14500, 18650, and 26650) and noticed that self-heat rates in 18650 and 14500 batteries are low compared to that of 26650. In addition to these findings about thermal runaway, LIBs exhibit temperature rise during charging/discharging conditions as well. The excessive rise in temperature may lead to hot spots, localized degradation, and temperature heterogeneity [21]. Such temperature heterogeneity is common in cylindrical batteries [22] during both the discharging or charging conditions. The extent of temperature rise is largely dependent on charging or discharging rate expressed as C rate, which is a measure of the charge and discharge current with respect to the nominal capacity of a battery. In other way, a battery with capacity of 1Ah can supply 1A current for one hour at discharge of 1 C rate. In case of discharging/charging at higher C rate, the heat generation is high that necessitates the use of efficient cooling systems. Thakur *et al.* [23] have discussed the BTMS with cold plate, composite phase change materials (PCMs), hybrid BTMS and some recent variations of BTMS for fast charging conditions. For both the charging and discharging conditions, batteries with PCM based cooling systems [24-27] is reported to have improved performance. Though, the control of such cooling system is difficult because of complex heat generation phenomenon, the fundamental equation of energy, continuity, and momentum [28] are used in simulating the thermal response of batteries. These studies revealed the thermal response of LIBs under different discharging/charging conditions, while Braga *et al.* [29] discussed the cell degradation phenomenon and gas release during thermal runaway.

The finding of the present work also raises a question about, which battery size is suitable for what type of EVs, wherein, power and energy density requirements are appropriately met with reduced risk of failure. In the hybrid electric vehicles, different sizes of cylindrical shells are being used, depending on the power and energy requirements as the Tesla base model use LIB 18650 while the Tesla model 3 uses LIB 21700 having a higher capacity [30]. The specification and application of batteries of different sizes is listed in Table 1.

**Table 1** Specifications of batteries (Manufacturer's data).

| Model | IFR 18650 | IFR22650 | IFR26650 | IFR32650 |
|---|---|---|---|---|
| Capacity (mAh) | 1500 | 2000 | 3000 | 5000 |
| Nominal voltage (V) | 3.2 | 3.2 | 3.2 | 3.2 |
| Charge voltage (V) | 3.65 | 3.65 | 3.65 | 3.65 |
| Anode | Graphite. | Graphite. | Graphite. | Graphite. |
| Cathode | $LiFePO_4$ | $LiFePO_4$ | $LiFePO_4$ | $LiFePO_4$ |
| Electrolyte material | Carbonate based | Carbonate based | Carbonate based | Carbonate based |
| Continuous maximum charge current | 1C 5A | 1C 5A | 1C 5A | 1C 5A |
| Continuous maximum discharge current | 3C 5A | 3C 5A | 3C 5A | 3C 5A |
| Width (mm) | 18 | 22 | 26 | 32.2 |
| Height (mm) | 65 | 65 | 65 | 65 |

| Weight (g) | 40 | 60 | 85 | 140 |
|---|---|---|---|---|
| Operating temperature range (°C) | Charge: 0-45 Discharge: 0-60 | Charge: 0-45 Discharge: 10-60 | Charge: 0-45 Discharge: 10-60 | Charge: 0-45 Discharge: 20-60 |
| Application | For future Tesla cars, E-bikes, LED Flash Lights | Car, E-bikes, Wheelchair | High-powered LED flashlights | Three-wheelers, High-powered LED flashlights |

Despite being designed as per the safety standards, several cases of the fire and explosion of hybrid electric vehicles are witnessed, prompting towards identification of the possible reasons of such incidents. Certainly, the thermal response of batteries is a major cause of failure, which is different for different-sized batteries [31-33]. It is well known that, battery size is proportional to stored energy, hence its energy release rate is different, consequently risk of thermal runaway is more for higher sized battery. These arguments necessitate to perform the size dependent failure studies of lithium-ion batteries. To the best of the author's knowledge, the size-dependent failure behavior of cylindrical LFPBs under mechanical abuse conditions has not been reported. Since, the batteries are available in many shapes and sizes, and the design of their internal components differ, hence post-mortem analysis of pristine battery is a need for a comparative assessment of safety features provided.

In this context, the present work reveals the size dependencies of LFPBs focusing thermal response under mechanical abuse conditions. The battery chemistry governs the energy density and capacity, however, the geometry (*that includes shape: cylindrical, prismatic, and pouch) and size: 18650, 22650, 26650, 32650 for cylindrical batteries*) play an important role in its safe operability. Considering the significance of size, the present work aims to analyze the effect of dimensional parameters of various components of LFPBs. To analyze the effect of dimensional parameters, the post-mortem analysis of four sizes of pristine LFPBs (18650, 22650, 26650, 32650) is conducted. The present work ignored the structural design of internal components and morphological analysis, which is mainly performed for *aged batteries* [34]. Moreover, size dependencies are analyzed by considering batteries of different diameters and comparing the failure load and temperature rise at the initiation of ISC. The mechanical behavior of LFPBs is segmented in different stages [35] when subjected to loading considered herein, revealing its layered and multicomponent structure.

## 2. Experimental

Post-mortem analysis of pristine LFPBs (SONY: IFR18650, IFR22650, IFR26650, and IFR32650) was conducted to examine the internal structure and components (Fig. 1 and Table 1). Subsequently, mechanical abuse testing was carried out with pristine samples of LFPBs of each size. In order to minimize the experimental uncertainties, the measurements were done using the calibrated instruments inside the laboratory environment. The thermocouple and other connections were ensured to be intact during the test. The abuse tests were conducted for three sets of samples of LFPBs for 18650 and 26650, while other sized samples were tested with the same configuration for similar loading situations.

*2.1 Port-mortem analysis*

The post-mortem analysis performed in the present work includes the disassembling of pristine LFPBs, identification of the battery components, measurements of the key dimensions of the sub-components, and the calculation of specific density, as reported in the literature [36]. The procedure involved discharging of the batteries to zero states of charge (SOC) and removing the non-conductive polymer covering from the outer casing. The subsequent procedure included the careful cutting of the outer casing near the positive and negative terminal using a hacksaw blade. The standard safety protocols are followed during all experiments. The disassembling of LFPBs can be dangerous due to the possibility of short-circuiting and release of harmful electrolytes or fumes [37]. Therefore, protective gloves and masks were used while performing it. Once the cell was opened, the spiral wound jellyroll was extracted from the casing after the removal of the positive terminal plate and the dismantling of components of the safety valve. After the complete disassembling, the various components were identified as shown in Fig. 1a and b.

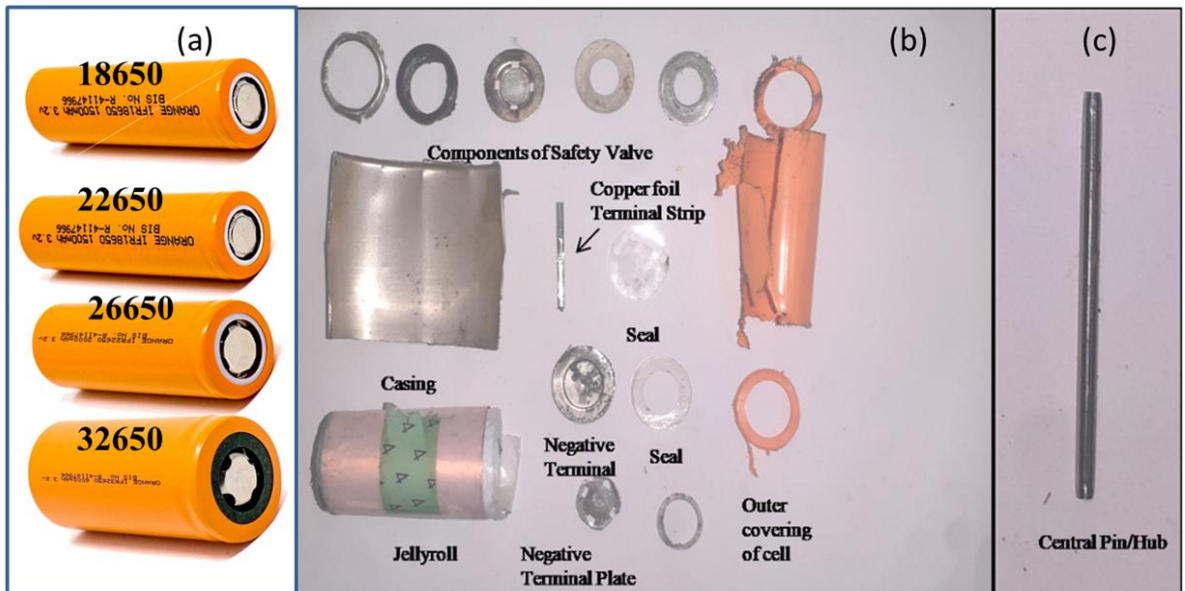

**Fig. 1.** (a) Picture of LFPB 18650, 22650, 26650 and 32650 (b) parts of LFPBs considered (Components of safety valve, copper foil terminal strip, seal, casing, jellyroll, negative terminal, negative terminal plate, outer covering of cell), (b) Control pin found only in LFPB 32650.

*2.2 Quasi-static mechanical testing*

Quasi-static lateral compression, longitudinal compression, and nail penetration tests were carried out to explore the mechanical behavior of the individual battery. The tests were performed in a calibrated universal testing machine (Werkstoffprüfmaschinen, Germany) of 40T capacity. The test setup for lateral compression, longitudinal compression, and nail penetration tests was conducted on LFPBs as shown in Fig. 2a-d. For nail penetration tests of LFPBs of 18650 and 22650 are conducted in a semi-closed fixture as shown in Fig.2c. To ensure the safety of the operator from the high-capacity LFPBs (26650 and 32650), the test were conducted in a completely closed fixture. The fixture included a hollow cylinder of thickness 4 mm and diameter 34 mm to hold the LIB inside it. A through hole was provided on to the fixture for holding the nail in place as shown in Fig.2d. The compression and nail penetration tests were conducted at a loading speed of 5 mm/min until the voltage drops to zero and the temperature starts rising. The temperature was measured by using K-Type thermocouple (having accuracy of ±1.5 °C as per IEC 584-2). The thermocouple was attached to the surface of a battery using Kapton tape at the positive terminal which exhibit higher temperature compared to that of negative terminal. The battery samples were connected to Arduino-based voltage module (MAX6650) and temperature module (MAX6675) as shown in Fig.2e and f for the *in-situ* recording of voltage and temperature data using Arduino-UNO.

The voltage and temperature modules were calibrated using a multimeter (DT830D) and a non-contact infrared thermometer (IT-1520). The voltage, temperature, displacement, and load data were recorded during the lateral, longitudinal compressions, and nail penetration tests. The experiment has been performed with a protective transparent shield of acrylic material to protect the test equipment and the operator.

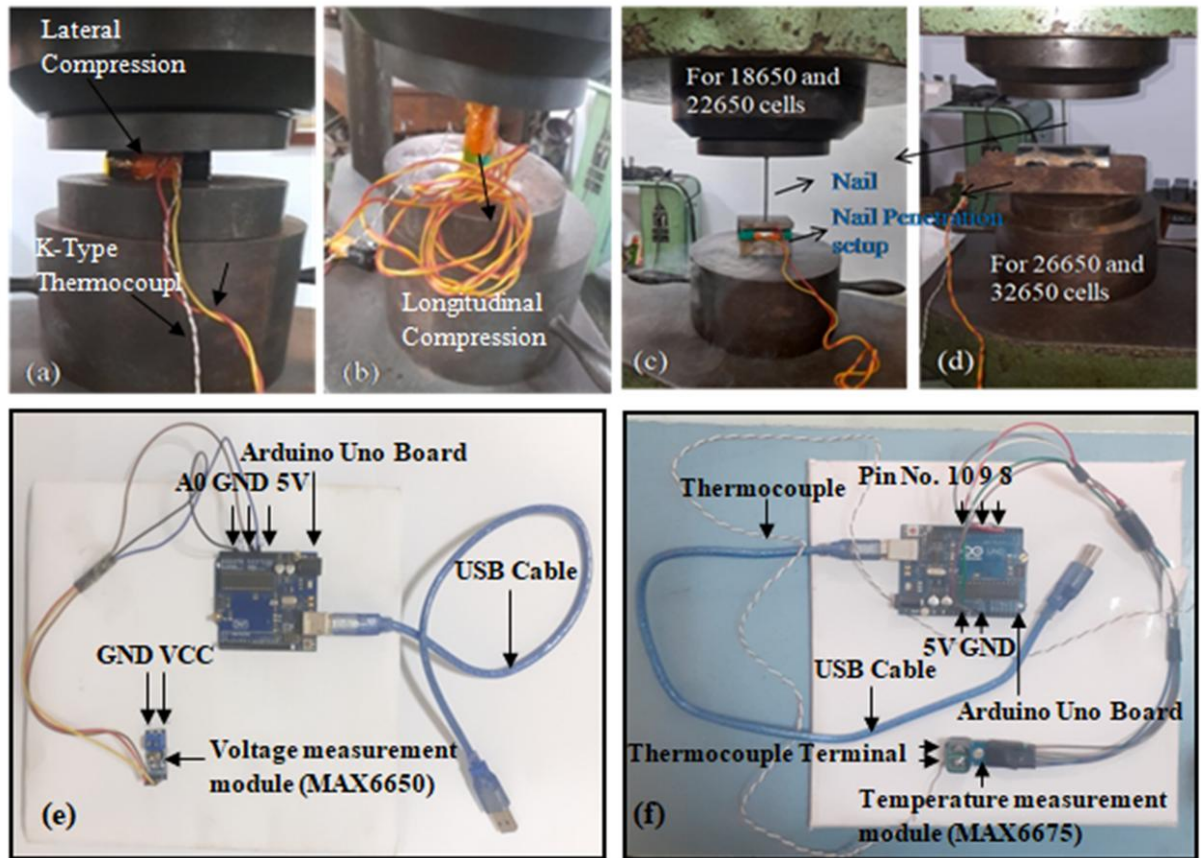

**Fig. 2.** Photographs of (a) lateral compression, (b) longitudinal compression, and (c,d) nail penetration tests performed using universal testing machine (e) voltage and (f) temperature measurement modules along with the connections made for recording the voltage and temperature during abuse tests.

3.       Results and discussion

The dimension of key components and subcomponents measured using Digital Micrometer (for thickness), Digital Vernier Calliper (for width) and measuring tape (for length) were listed in Table 2. It was found that the same components are present in all the samples except an additional central pin in LFPB 32650. The hollow central pin facilitates the venting of fumes through itin case of short-circuiting and separator burnout.

Safety valve assembly included some sub-components such as a positive temperature coefficient (PTC) device, a plate with safety vents, a positive terminal plate, and a gasket seal. The jellyroll, which includes an anode, cathode, and two layers of polymeric porous membrane separator, is the most essential part that stores the energy. The active anode material i.e., graphite was found to be deposited on both the sides of copper current collector surfaces. The active cathode material i.e., LFP coating was seen on both the sides of an aluminum current collector. A separator that is permeable to ionic flow was found to be kept between the electrodes. The ions travel to the anode from the cathode, during the charging process and vice versa during discharging through the separator. The measurements for the sub-components and the calculated specific density of the LFPBs are listed in Table 2.

**Table 2** Dimensions of sub-components of LFPB 18650, 22650, 26650, and 32650.

| LFPB sub-components | 18650 | 22650 | 26650 | 32650 |
|---|---|---|---|---|
| Copper foil thickness | 0.031 mm | 0.049 mm | 0.097 mm | 0.029 mm |
| Aluminum foil thickness | 0.113 mm | 0.046 mm | 0.035 mm | 0.016 mm |
| Thickness of graphite coat on copper foil | 0.048 mm | 0.041mm | 0.013 mm | 0.046 mm |
| Thickness of LFP coat on aluminum foil | 0.032 mm | 0.0805 mm | 0.0815 mm | 0.087 mm |
| Separator thickness | 0.034 mm | 0.034 mm | 0.020 mm | 0.016 mm |
| Casing thickness | 0.345 mm | 0.431 mm | 0.383 mm | 0.387 mm |
| Length of Copper foil | 74.8 cm | 95.0652 cm | 152 cm | 278 cm |
| Length of Aluminum foil | 69.5 cm | 95.0652 cm | 142 cm | 276 cm |
| Length of separator | 164 cm | 190.1304 cm | 312 cm | 568 cm |
| Width of jellyroll layers | 6 cm | 6 cm | 6 cm | 6 cm |
| Copper terminal thickness | 0.559 mm | 0.155 mm | 0.195 mm | 0.08 mm |
| Copper terminal width | 4.2 mm | 3.908 mm | 5.64 mm | 5.57 mm |
| Aluminum terminal Thickness | 0.1 mm | 0.283 mm | 0.126 mm | --- |
| Aluminum terminal width | 4.92 mm | 4.447 mm | 3.03 mm | 5.99 mm |
| Volume of graphite coat on copper foil | 4308.48 mm$^3$ | 4,563.1296 mm$^3$ | 2,371.2 mm$^3$ | 15,345.6 mm$^3$ |
| Volume of LFP coat on aluminum foil | 2668.8 mm$^3$ | 9,183.29832 mm$^3$ | 13,887.6 mm$^3$ | 28814.4 mm$^3$ |
| Specific energy | 120 (Wh/kg) | 106.67 (Wh/kg) | 112.94 (Wh/kg) | 114.28 (Wh/kg) |

The dimensions of the sub-components of LFPBs are examined to understand their contribution to the energy storage capacity. The capacity of batteries depends on the volume of active materials on electrodes i.e., graphite on copper foil (Fig. 3a) and LFPs on aluminum foil (Fig. 3b), and the total volume of LFP and graphite together (Fig. 3c). The volume of LFP is highest in LFPB 32650 but ratio of LFP to graphite, deposited on the copper and aluminum foil respectively is maximum for LFPB 26650 (i.e., 5.85) and minimum for LFPB 18650 (i.e., 0.61). It can be seen from the plot that, among all the LFPBs considered, the least amount of copper was used in 26650 cells which is having 3000 mAh capacity. Although a higher amount of copper is used, the capacity of 18650 and 22650 are 1500 mAh and 2000 mAh respectively, which is lower than the capacity of LFPB 26650 (Fig. 3). It indicates that, the battery capacity is not solely dependent on the amount of LFP, but it depends also on the ratio of LFP and graphite is an important parameter governing the capacity. Also, considering the aspects of effective utilization of materials [38, 39], towards ensuring sustainable development in the field of energy storage, such findings may help in developing LFPBs with reduced copper.

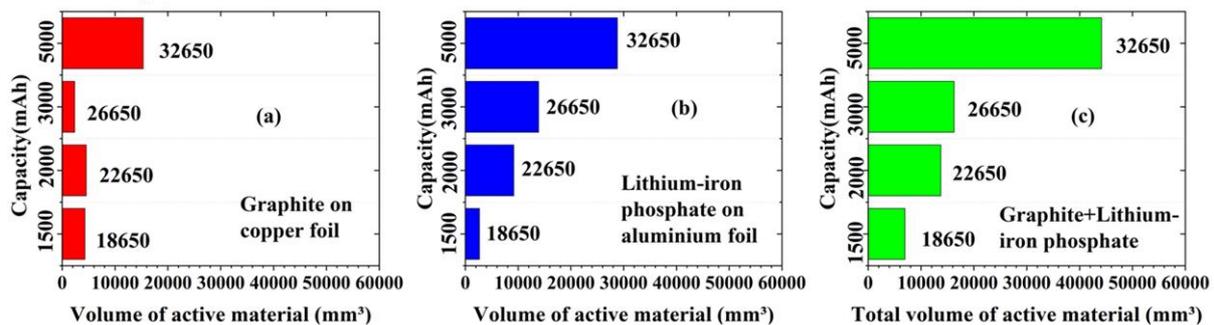

**Fig. 3**. Active materials in LFPB18650, 22650, 26650, and 32650, (a) volume of graphite on copper foil, (b) volume of lithium-iron phosphate (LFP) on aluminum foil, (c) total volume with active materials of LFP and graphite (on aluminum and copper foil both).

The findings from the post-mortem analysis evidenced a direct correlation between the volume of active materials and the capacity, which governs the size of the battery. It may subsequently affect the failure behavior of each LFPBs considered in the present investigation. To validate that, an *in-depth* investigation of the size-dependent failure behavior of LFPBs subjected to the lateral, longitudinal compression tests, and nail penetrations tests are characterized by the onset of ISC. One can easily visualize the failure, evident in the form of fumes released as shown in Fig. 4a-d. The fumes may contain toxic lithium hexafluorophosphate ($LiPF_6$) which evaporates and immediately converts into short-lived toxic hydrogen fluoride in the presence of moisture. According to the previously reported work [37], the rate of release of hydrogen fluoride for cylindrical LFPBs compared to several other battery chemistries was found to be 24 mg/Wh at 100% SOC in the case of the fire test. Though, the present work does not include an estimation of the rate of release of fumes, the rate of release was visually distinct after the event of a short circuit at a time delay of 1-3sec (recorded during the tests) between the release of fumes and onset of short-circuit in lateral and nail penetration tests. In the case of longitudinal compression, a time delay of 5-8 sec was observed. Such delay in the release of fumes increases the risk of explosion too and should be avoided straightly. It should be noted again that, the tests were conducted by following the safety procedure keeping the place of the test well-ventilated with operating exhaust systems, and wearing FFP3 respirator masks. Fig. 4a shows the abused sample corresponding to lateral compression of LFPBs 18650 displaying rapid release of fumes compared to that observed in Fig. 4b for longitudinal compression of LFPBs 32650. This indicates a high rate of chemical reactions occurs in the case of lateral compression due to the involvement of full length (i.e., 65 mm) of active materials on the electrodes in the short-circuit event compared to that occurring in longitudinal compression. Such rapid release of fumes and electrolytes was facilitated by the provision of vents in the samples. On the other hand, the rate of release of fumes and electrolytes in the case of nail penetration tests was less (Fig. 4c and d) due to the short circuits at smaller regions or localized spots under the nail tip. Fig. 4e-g illustrates the area of active material involved during the compression tests performed. To avoid the chance of battery explosion due to the high rate of exothermic chemical reactions during short-circuiting, blockage of the vents should be avoided. The chance of battery explosion is reduced in lateral compression and nail penetration, as fumes and electrolytes release are facilitated by the presence of vents and holes due to nail penetration, which is blocked due to the machine crosshead in case of longitudinal compression.

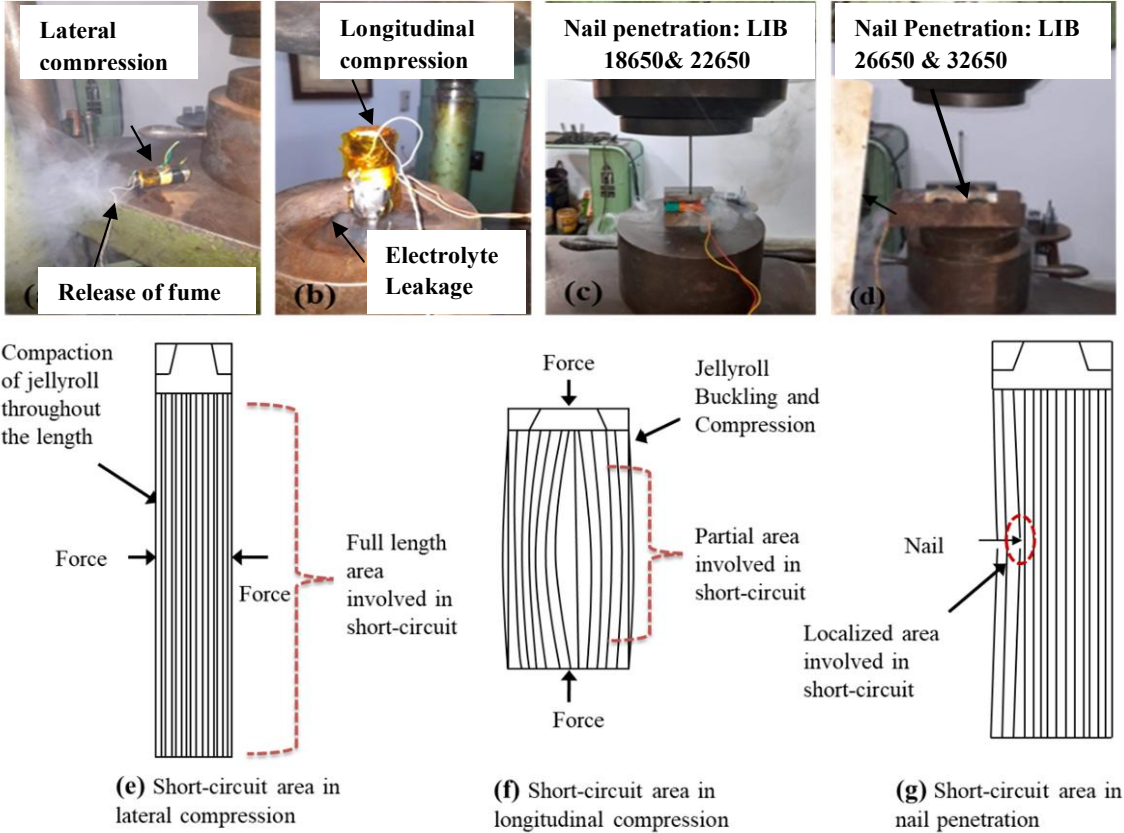

**Fig. 4.** Release of fumes at the ISC during mechanical abuse tests: (a) lateral compression (b) longitudinal compression, and nail penetration with (c) semi-closed and (d) fully closed sample fixture. (e) Representation of compaction of jellyroll in lateral compression, (f) bending and buckling of jellyroll in longitudinal compression, and (g) localized compaction in nail penetration.

The *in-situ* measurement of applied force, voltage, and temperature data with the cross-head displacements during the quasi-static abused loading are recorded and illustrated in Fig. 5. The load-displacement curve is used to identify the failure load, which is considered as the load at the instant of voltage drop due to mechanical failure or rupture of jellyroll during ISC. The load vs displacement can be segmented into four stages [35]as follows: (i) stage-I represents the compression of the cell tab and casing of the battery till the jellyroll and cell tab contact is developed, (ii) Stage-II shows the onset of jellyroll compression after casing compression is complete, (iii) Stage-III denotes the compaction stage wherein jellyroll is compressed and bent leading to buckling and short-circuiting at the spot of the contact point, and (iv) last stage, stage-IV marks the complete failure of battery after severe or hard short-circuit. Since the present work focuses on the onset of failure of LFPBs, the load vs displacement curve behavior showed evidence of three stages only. The onset of the short circuit, evident as a drop in voltage, which is called a soft short circuit, becomes severe if further compaction or loading is continued resulting in a hard short circuit. Therefore, the first event of voltage drop can be considered as a reference instant of time that can be used as an early warning system to control the catastrophic failure leading to thermal runaway. Referring to Fig. 5a-d, for lateral compression, the failure load for LFPBs 18650, 22650, 26650, and 32650 is found to be 44, 31.88, 12.78, and 6.68kN respectively. It implies that the smaller the size, the larger the failure load for the considered LFPBs. This leads to greater compaction of jellyroll i.e., higher displacement, resulting in a higher temperature at the onset of ISC.

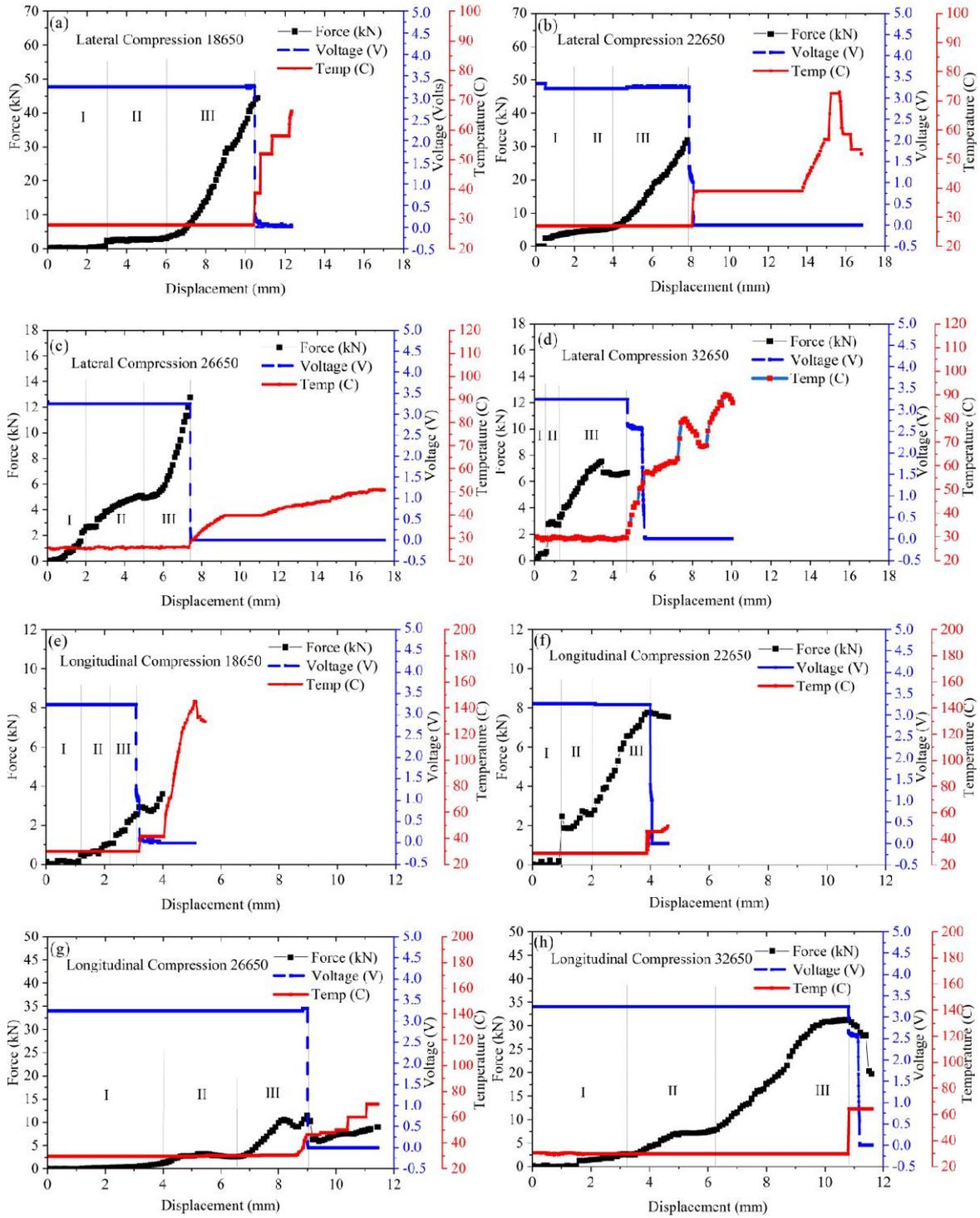

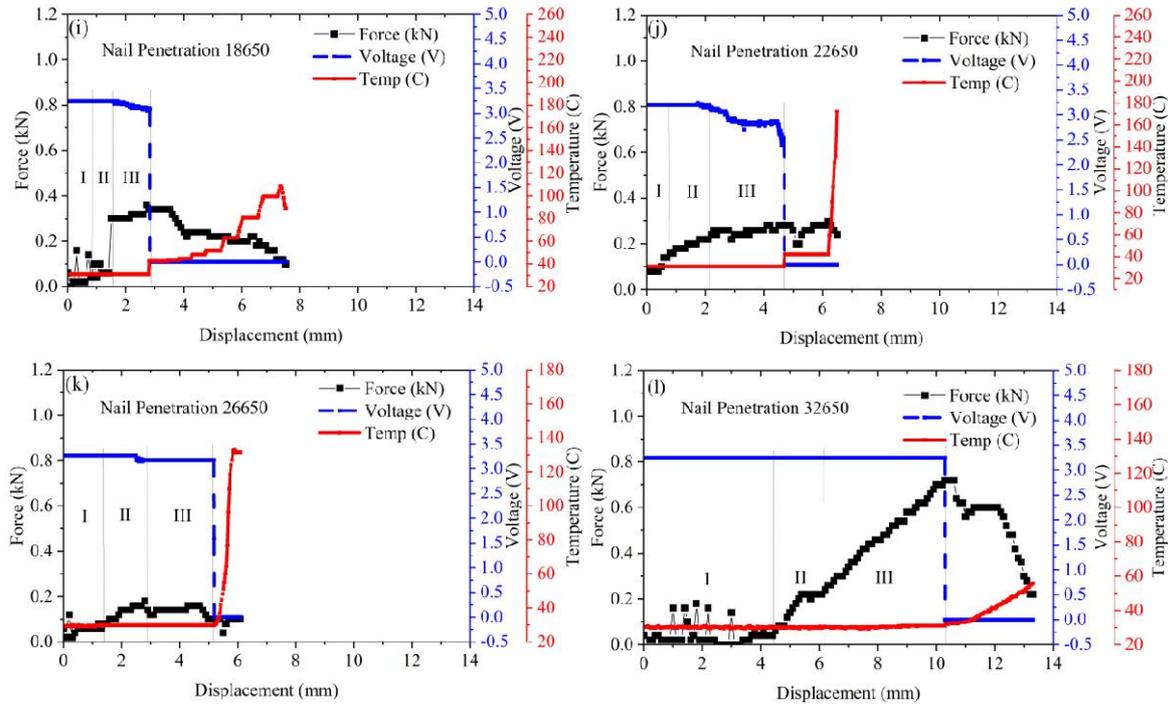

**Fig. 5.** Variation of force, voltage, and temperature with displacement: for lateral compression in (a) 18650, (b) 22650, (c) 26650, (d) 32650, for longitudinal compression in (e)18650, (f) 22650, (g) 26650, (h) 32650, for nail penetration (i) 18650, (j) 22650, (k) 26650, (l) 32650.

Referring to Fig. 5e-h, for longitudinal compression, the failure load for LFPBs 18650, 22650, 26650, and 32650 is found to be 2.92, 7.76, 11.44, and 31.24kN respectively. It shows that, smaller the size, smaller is the failure load for the considered LFPBs. Under longitudinal compression, the load acted along the axis of the LFP considered. This leads to the initial bending of the jellyroll with eventual buckling (Fig 4 f). Such buckling deformations cause smaller area of direct contact between the electrodes in small sized battery (LFPB 18650) resulting in the lowest temperature rise at the onset of ISC for it compared to that of LFPB 32650, which displayed highest temperature rise. On comparing the load-displacement curve for lateral compression, LFPBs 32650 shows a sudden drop in failure load in stage III. Such a drop in load (Fig. 5g) can be attributed to the sudden compaction of the hollow central pin found only in LFPB 32650. The hollow central pin provides the passage to the gaseous products/fumes to the vent. Additionally, it provides buckling strength to the jellyroll displaying a higher failure load as in Fig. 5h, corresponding to longitudinal compression. These results suggest that there should be an optimized dimension of the battery such that the chances of thermal runaway due to mechanical abuse can be controlled. Referring to Fig. 5i-l, in the case of nail penetration tests, the failure load for the onset of ISC is very high for LFPB 32650 compared to the other battery sizes. Also, the failure behavior of the LFPBs shows that the nail penetration depth required for the onset of ISC decreases with the decrease in the size of the battery for the considered LFPBs.

According to the literature [40], the safe operating temperature of LIBs lies in the range of 20-60 °C during the discharge process. Fig. 5 shows that there is an increase in surface temperature upon increasing the loading. The temperature rise is due to the exothermic chemical reactions inside the LFPBs during ISC. Temperature also rises due to the Joule heating, hence the combined heat due to the Joule heat and exothermic chemical reaction post short-circuiting (soft and hard) are the reason of further rise in temperature. Both, the soft and hard short circuit are phenomenon wherein flow of charge takes place between the battery electrodes through zero resistance path that is created due to failure of battery separator.

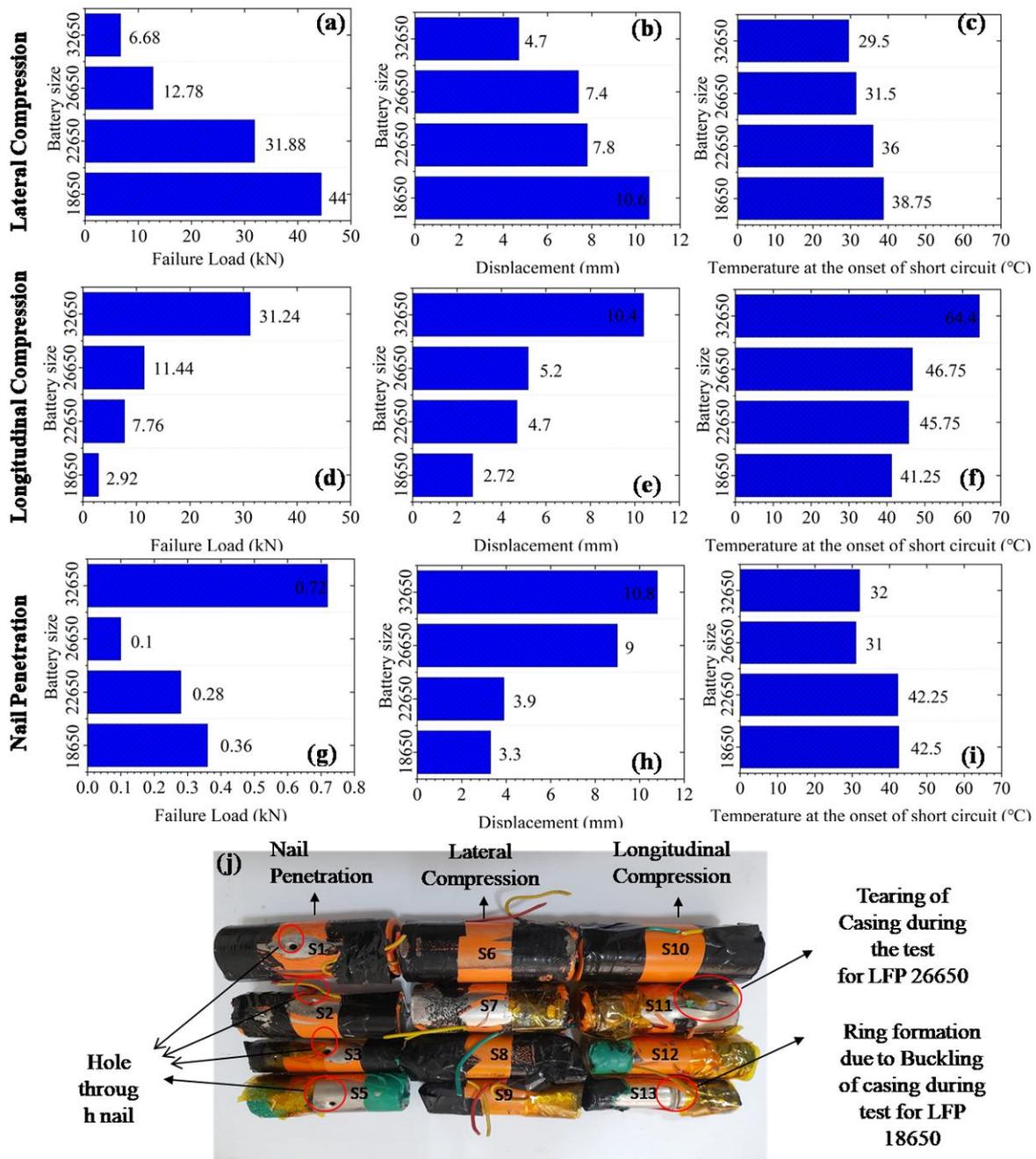

**Fig. 6.** Failure load, displacement, and temperature at the onset of short circuit corresponding to (a-c) lateral compression, (d-f) longitudinal compression, and (g-i) nail penetration. (j) Pictures of failed LFPB samples after nail penetration (S1 to S4), lateral (S6 to S9), and longitudinal compression (S10 to S13) tests.

The soft short circuit involve few numbers of electrode layers to come into direct contact with each other in comparison to the hard short circuiting. Consequently, the drop in voltage and temperature rise is different at these short-circuiting cases. Involvement of more layers of electrodes in short circuiting results into large amount of heat generation, which can lead to complete failure due to uncontrolled chemical reaction. Such cases can fall under the category of hard short circuiting leading to excessive temperature rise which may result into thermal runaway and fire or explosion. This temperature rise should be kept within the safe limit to prevent the thermal runaway and the battery thermal management systems should be designed accordingly to respond to such undesirable events. The soft short circuit, is a triggering event for thermal runaway, therefore, the present test results can be used to develop appropriate methodologies for early warning systems to detect failure. For better clarification, the size dependencies of commercial LFPBs considered in the present work are analyzed using Fig. 6 by a comparative assessment of failure behavior. Fig. 6a shows decreases in failure load with an increase in the size of the LFPBs in the case of lateral compression. Referring to Fig. 6b and c, for LFPB18650, the temperature rise is about 38.75 °C, which is slightly higher than the optimum temperature range (15-35 °C) for LIB in literature [6]. Further, the rise in temperature due to the short circuit is delayed by the decrease in the diameter of the LFPBs for the lateral compression test. The delay in temperature rise implies less heat generation rate. Such temperature rise data therefore becomes important for faster heat diffusion and effective thermal management. Fig. 6d shows an increase in failure load with the increase in the size of the LFPBs in the case of lateral compression. Referring to Fig. 6e and f, the rise in temperature starts earliest for the LFPBs 18650 cell and gets delayed with the increase in diameter of the LFPBs in case of longitudinal compression. Referring to the results of nail penetration tests in Fig. 6g, the value of failure load for LFPBs 32650 is 0.72kN, which is the highest among the batteries of other sizes. Referring to Fig. 6h and i, the short circuit temperature is almost the same for LFP 18650 and 22650. Lower short-circuit temperature is noticed for higher-sized LFPBs such as that observed for LFP 26650 and 32650 cells in comparison to LFP 18650 and 22650, with the earliest short circuit in LFP 18650 in comparison to the other larger LFPBs subjected to nail penetration tests.

The failed samples after the mechanical abuse tests are shown in Fig. 6j. The formation of the hole on nail penetration tested LFPBs (Fig. 6j, S1 to S4) was visible in all the tested samples. Rings formed (Fig. 6j, S13) in LFPB 18650 indicate the cause of ISC due to buckling in longitudinal compression tests. During longitudinal compression testing of LFPB 26650, there was a sudden rupture and tearing of the casing (Fig. 6j, S11) leading to a mild explosion due to the blockage of the vent by the cross head. The high-temperature resistant tape was used to avoid malfunctioning/disconnection of the wire soldered to the LFPB samples during tests. The subsequent tests were conducted after creating holes in the thermal tape keeping the vent space open during the tests.

4.  **Conclusion**

The present work included post-mortem analysis of pristine cells. It was noted that a central pin was provided additionally in LFPB 32650 as an extra safety measure to facilitate the passage to gas/fumes from the negative to the positive terminal in case of thermal runaway conditions. Other important conclusions of the present work highlighting the size dependencies on the failure behavior of LFPBs considered are listed below:

1) Failure load decrease with an increase in diameter of the LFPBs for lateral compression tests, alternatively, it increases with an increase in size for longitudinal compression tests. During the nail penetration test, failure load is maximum for LFPB 32650 in comparison to other batteries considered.
2) The onset of internal short circuit delays with an increase in diameter of the batteries for longitudinal and nail penetrations tests, while it occurs earliest in the case of batteries with a larger size for lateral compression tests.
3) The temperature rise at the onset of short circuit is different for different battery sizes. It is maximum (64.4 °C) for LFPB 32650 in case of longitudinal compression and minimum (29.5 °C) for lateral compression. For longitudinal compression test, at the onset of the short circuit, the variation in the temperature, above the ambient value, was found to be the highest for LFPB 32650 showing a rise of 36.4 °C. On comparing the values of the temperature rise at the onset of short circuit, a balanced thermal behavior of LIB 26650 is revealed for all mechanical abuse conditions considered.

4) Time delay of up to 3 sec between the release of fumes and onset of the short circuit was observed in lateral and nail penetration tests, while it was up to 8 sec in longitudinal compression tests. An increase in the delay in the release of gas/fumes increase the risk of gas accumulation and subsequent explosion. The risk of the explosion also increases if vents are blocked, which was seen in the case of LFPB 26650, wherein the vent was blocked due to the Kapton taping to keep the thermocouple in position during test.

These test results may have significant implications for the development of testing standards for batteries and early failure detection systems for battery packs towards improving the safety of electric vehicles.

**Declaration of competing interest**

The authors declare that they have no known competing financial interests or personal relationships that could have appeared to influence the work reported in this paper.

**CRediT authorship contribution statement**

**Vishesh Shukla:** Conceptualization, Methodology, Draft preparation, Experimentation, and Data curation. **Ashutosh Mishra:** Conceptualization, Supervision, Experimentation, Data curation, Visualization, Writing- Reviewing and Editing, Project Administration. **Jagadeesh Sure:** Visualization, Writing- Reviewing and Editing. **Subrata Ghosh:** Visualization, Writing- Reviewing and Editing. **R.P. Tewari:** Conceptualization, Supervision, Visualization, Writing- Reviewing and Editing.

**Acknowledgement**

Authors would like to acknowledge the necessary infrastructural support provided by Motilal Nehru National Institute of Technology Allahabad, Prayagraj, India. S. G. acknowledge the European commission for the award of Marie Skłodowska-Curie Postdoctoral Fellowship (MSCA-PF: Grant number-ENHANCER-101067998).

**Funding resources**

This research did not receive any specific grant from funding agencies in the public, commercial, or not-for-profit sectors.